\documentclass[12pt]{iopart}
\usepackage{graphics}
\usepackage{mathtools,cuted}
\usepackage{color}
\usepackage{lipsum}
\usepackage{multirow}
\usepackage{threeparttable}

\begin{document}

\title{Mean-field theory for the Nagel-Schreckenberg model with overtaking strategy}

\author{Zhu Su$^{1\ast}$, Weibing Deng$^2$, Jihui Han$^{3}$, Wei Li $^{2}$, Xu Cai$^{2}$}

\address{$^1$National Engineering Laboratory for Technology of Big Data Applications in Education, Central China Normal University, Wuhan, P. R. China}
%\address{$^2$National Engineering Research Center for E-Learning}
\address{$^2$Complexity Science Center, Institute of Particle Physics, Central China Normal University, Wuhan, P. R. China}  
\address{$^3$School of Computer and Communication Engineering, Zhengzhou University of Light Industry, Zhengzhou, P.R. China}
\ead{suz@mail.ccnu.edu.cn}  
\vspace{10pt}
\begin{indented}
\item[]May 2018
\end{indented}

\begin{abstract}
Based on the Nagel-Schreckenberg (NS) model with periodic boundary conditions, a modified model considered overtaking strategy (NSOS) has been proposed \cite{su2016occurrence,su2016the}. In this paper, we focus on the theoretical analysis of traffic flow for NSOS model by using mean-field method. In the special case of $v_{max}=1$ where vehicles can not overtake preceding ones, the features of stationary state can be obtained exactly. However, in the case of $v_{max}>1$ where overtaking happens, some approximative methods have to be took into account.  The main results are that we find the reason why traffic flow is increased in the regime where densities exceed the maximum flow density, and the influence of traffic flow on the transition density is dominated by the braking probability $p$.   
\end{abstract}

\section{Introduction}
Various dynamical models \cite{Chowdhury2000,Helbing2001} have been proposed to explain the complexity phenomena generated by traffic flow.  From the microscopic point of view, the vehicular traffic system can be regarded as being composed of interacting particles driven far from equilibrium, each individual vehicle is represented by a particle, and the way that they influence others' movement is treated as the interactions among particles. Therefore, vehicular traffic offers the possibility to study various fundamental aspects of the dynamics of non-equilibrium systems which are of interest in statistical physics. 

During the last two decades, cellular automata (CA) \cite{Wolfram1983} have obtained popularity due to their simplicity and their ability to simulate large networks. One of the early CA based on traffic models is the NS model \cite{Nagel1992} developed by Nagel and Schreckenberg, and then a large amount of improved versions have been proposed by imposing some conditions on NS model to make it more realistic. In the NS model, the road is divided into sites, each site can be either empty or occupied, and all the space, time and velocities are discrete. Under periodic boundary conditions, the number of vehicles on the road remains unchanged. The state of vehicle is characterized by an internal parameter $v$ ($v=0,\dots,v_{max}$), where $v_{max}$ is the maximum velocity. To obtain the system's state in the next time, one could adopt the following rules to all vehicles at the same time (parallel dynamics): (1) The first step is an acceleration process, if a vehicle's velocity ($v$) is lower than the speed limit ($v_{max}$), its velocity is advanced by one. (2) The second step is designed to avoid accidents, if two adjacent vehicles have $h$ empty sites between each other, and the following vehicle has a speed larger than $h$, then its velocity is reduced to $h$. (3) The third step is considered random braking, a noise with probability $p$ to reduce the velocity of a moving vehicle ($v>0$) to $v-1$. (4) The last step deals with the vehicle' movement, which enables the position of a vehicle to be advanced by its speed $v$. 

Considered the overtaking case, the NSOS model has been proposed \cite{su2016occurrence,su2016the}, where every vehicle could be an overtaking one with probability $q$ at each time step. The model exists due to the following facts: (1) Overtaking obviously happens a lot, especially when the preceding vehicles move quite slowly. (2) The overtaking vehicles are used to go back to the original lane once overtake successfully. (3) When the preceding vehicles are overtaken, they would slow down rather than accelerate. 

Several approaches for analytical descriptions of the NS model have been studied \cite{Schadschneider1997,Schadschneider1999,Schreckenberg1995,0305-4470-26-15-011,0305-4470-30-4-005,0305-4470-31-11-003,0305-4470-31-22-008}, and they yield the exact solution for the special case of $v_{max}=1$ and are good approximations for higher values in the case of $v_{max}>1$. In this paper, we use mean-field method \cite{Schreckenberg1995} to analyze the NSOS model. The general express equations have been obtained in the stationary state in the case of $v_{max}>2$. In the special case of $v_{max}=1$, where overtaking vehicles can not overtake the preceding ones, the detail results in the stationary state have been done. In order to find the qualitatively different behaviors in the case of $v_{max} \geq 2$, we calculate the results for  $v_{max}=2$ where overtaking happens. 

\begin{figure}[h]
	\centering
	\includegraphics[scale=0.2]{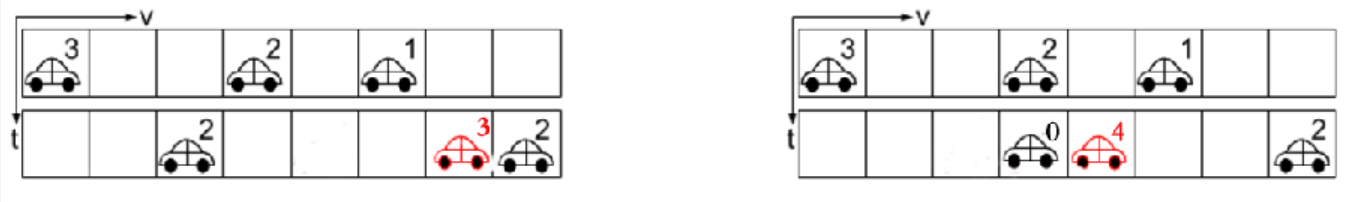}
	\caption{\label{graph}(Color online) Schematic graph of updating and moving in the NSOS model. The numbers in the sites represent the velocities after moving. The red cars are overtaking vehicles (left). The second car is an overtaking vehicle, it has tried to overtake the preceding one but failed (right). The first car is an overtaking vehicle, it has successfully overtook the preceding one.}
\end{figure}
%\begin{figure}[h]
%	\hspace{10pc}
%	\includegraphics[scale=0.35]{Figure2.eps}
%	\caption{\label{spacetime}(Color online) Space-time diagram for $L=70$, $p=0.5$, $q=0.25$, $v_{max}=5 $ and $\rho=0.5$. Stars stand for free site, numbers stand for the velocity of a vehicle in this site after moving. With overtaking probability $q$, overtaking vehicles (red circles) try to overtake the preceding vehicle and some have succeeded. Jams still occur in the case of $\rho=0.5$. }
%\end{figure}

The paper is organized as follows: First briefly introduce the update rules of the NSOS model in Section $2$, then analyze this model by applying mean-field method in Section $3$, and the summary is discussed in the final Section. 

\section{NSOS model}
NSOS model is based on the NS model which considering overtaking strategy with probability $q$ \cite{su2016occurrence,su2016the}. In this model, overtaking vehicles are picked up randomly with probability $q$ at every time, and they overtake preceding ones depending on their configurations in the next time. Some assumptions have been considered as followings: (1) Overtaking vehicles would brake if they reached the same location with their preceding ones in the next time to avoid collisions. (2) Each overtaking vehicle is only able to overtake one per time step. (3) The overtaking vehicle locates in front of preceding one once overtake successfully. Similar to NS model, all the overtaking vehicles decrease by one with braking probability $p$ except for the successful overtaking ones. For convenience, we name the vehicles which are not overtaking as ordinary ones, and update their velocities according to the NS model. The detailed updating rules of the NSOS model are as follows:
\begin{description}
	\item[1.] At time $t$, the $j$th vehicle becomes an overtaking vehicle with probability $q$, otherwise it is an ordinary one.
	\item[2.] Update the velocity:
	\begin{description}
		\item[(I)] If the $j$th vehicle is an ordinary one:
		\begin{description}
			\item[(1)] Acceleration: \\
			$v(j,t_{1}) \rightarrow min(v(j,t)+1, v_{max})$.
			\item[(2)] Deceleration: \\
			$v(j,t_{2}) \rightarrow min(v(j,t_{1}),d(j,t))$.
			\item[(3)] Random braking: \\
			$v(j,t_3) \rightarrow max(v(j,t_{2})-1,0)$ with the probability $p$.
		\end{description}
		\item[(II)] If the $j$th vehicle is an overtaking one:
		\begin{description}
			\item[(1)] Acceleration: \\
			$v(j,t_{1}) \rightarrow 	min(v(j,t)+1, v_{max})$.
			\item[(2)] If $v(j,t_{1}) > d(j,t) +v(j+1,t+1)$, the position $d(j,t)+v(j+1,t+1)+1$ is empty and the $(j+1)$th vehicle does not overtake successfully,
			\item[(i)] Overtaking: \\
			$v(j,t_{3}) \rightarrow d(j,t)+v(j+1,t+1)+1.$
			\item[(3)] Otherwise,
			\item[(i)] Deceleration: \\
			$v(j,t_{2}) \rightarrow min(d(j,t)+v(j+1,t+1)-a,v(j,t_{1}))$.
			\item[(ii)] Random braking with probability $p$: \\
			$v(j,t_{3}) \rightarrow min(v(j,t_{2})-1,0)$.
		\end{description}	
	\end{description}		
	\item[3.] Movement:\\
	$x(j,t+1)=x(j,t)+v(j,t_{3})$.
\end{description}

Here, $v(j,t)$ denotes the velocity of the $j$th vehicle at time $t$ and $x(j,t)$ denotes its corresponding position. The number of empty sites in front of the $j$th vehicle is denoted by $d(j,t)=x(j+1,t)-x(j,t)-1$. To avoid collisions, we assume $a=2$ if the $(j+1)$th overtaking vehicle overtakes successfully. In other cases, $a=1$. An illustration of the NSOS model can be found in Fig.\,\ref{graph}. The numbers in the sites stand for their velocities after moving. In the left graph, the second car is an overtaking vehicle, it has tried to overtake the preceding one but failed. In the right graph, the first car is an overtaking vehicle, which has successfully overtook the preceding one.

Since the velocity of overtaking vehicle at time $t+1$ is relative to its preceding one, we should know the preceding vehicle's velocity at time $t+1$ first. Fortunately, the velocity of ordinary vehicle is independent of the preceding one, so we could pick up ordinary vehicles and update their velocity first, and then the rear vehicles. In our simulations, we use parallel update and periodic boundary conditions, assume the first and last vehicles as ordinary ones all the time, update their velocities first, and then update others' velocities from $(N-1)th$ car to the second one. 

\section{Mean-field theorem}
The simplest analytical approach to the NS model is a microscopic mean-field (MF) theory \cite{Schreckenberg1995}. Here one considers the probability $c_{\alpha}(i,t)$ of vehicles with velocity $\alpha$ at site $i$ and time $t$, and we denote the probability that there is no vehicle at site $i$ ($i=1,2,3,...,L$) at time $t$ by $d(i,t)$. In the MF approach, correlations between sites are completely neglected. Therefore one has the normalization condition for all sites and all time steps,
\begin{equation}
d(i,t)+\sum\limits_{\alpha=0}^{v_{max}}c_{\alpha}(i,t)=1 .
\end{equation}

Denoting with $c(i,t)$ the total probability for site $i$ to be occupied at time step $t$, i.e. $\sum_{\alpha=0}^{v_{max}}c_{\alpha}(i,t)$, one simply has $d(i,t)+c(i,t)=1$. In our model each car could be an ordinary one with probability $\bar{q}$ (here, $\bar {q}=1-q$) or overtaking one with probability $q$. Therefore we have the equation $c=c^{\ast}+c^{\dagger}$, where $c^{\ast}=qc$ denotes the probability of being overtaking vehicles and $c^{\dagger}=\bar{q}c$ denotes the probability of being ordinary vehicles. Since the update rules of overtaking vehicles relate to the next time positions of preceding ones, the configurations of them should be obtained first. Here we adopt $C(i+j,t)$ ($D(i+j,t)=1-C(i+j,t)$) as the probability that there is (not) a vehicle at site $i+j$ ($j=1,2,3,...,v_{max}$) at the next time before one updates the site $i$ at time $t$, need to say that, here, we just consider the vehicles in front of site $i$. We update the ordinary vehicles first, and choose the site of an ordinary vehicle as a starting site to update the rear vehicles. 

According to the update rules, the time evolution of these probability distributions can be described by the following sets of equations:

(1) For the ordinary vehicles, the update rules are the same with the ones in the original NS model, therefore the MF equations for the stationary state ($t \rightarrow \infty$) read \cite{Schreckenberg1995} : 

\begin{equation}\label{NS}
\begin{split}
c_{0}^{\dagger}& =(c+pd)c_{0}^{\dagger}+(1+pd)c\sum\limits_{\beta=1}^{v_{max}}c_{\beta}^{\dagger},\\
c_{\alpha}^{\dagger}& =d^{\alpha}[\bar{p}c_{\alpha-1}^{\dagger}+(\bar{p}c+pd)c_{\alpha}^{\dagger}+(\bar{p}+pd)c\sum\limits_{\beta=\alpha+1}^{v_{max}}c_{\beta}^{\dagger}], \qquad 0<\alpha<v_{max}\\
c_{v_{max}}^{\dagger}& =\bar{p}d^{v_{max}}(c_{v_{max}-1}^{\dagger}+c_{v_{max}}^{\dagger}).\\
\end{split}
\end{equation}
(2) For the overtaking vehicles ($v_{max} > 2$):\\
(i) The acceleration stage
\begin{equation}\label{NSOS1}
\begin{split}
c_{0}^{\ast}(i,t_{1})& =0,\\
c_{\alpha}^{\ast}(i,t_{1})& =c_{\alpha-1}^{\ast}(i,t), \qquad 0<\alpha<v_{max}\\
c_{v_{max}}^{\ast}(i,t_{1})& =c_{v_{max}}^{\ast}(i,t)+c_{v_{max}-1}^{\ast}(i,t).
\end{split}
\end{equation} 
(ii) The deceleration stage:\\
\begin{equation}
\begin{split}
c_{0}^{\ast}(i,t_{2})& =c_{0}^{\ast}(i,t_{1}),\\
c_{1}^{\ast}(i,t_{2})& =D(i+1,t)c_{1}^{\ast}(i,t_{1})+D(i+1,t)C(i+2,t)c_{2}^{\ast}(i,t_{1}),\\
c_{\alpha}^{\ast}(i,t_{2})& =\underbrace{c_{\alpha}^{\ast}(i,t_{1})\prod_{j=1}^{\alpha}D(i+j,t)}_\text{I} +\underbrace{c_{\alpha+1}^{\ast}(i,t_{1})\prod_{j=1}^{\alpha}D(i+j,t)C(i+\alpha+1,t)}_\text{II}\\
&\quad +\underbrace{C(i+\alpha+2,t)C(i+\alpha+1,t)\prod_{j=1}^{\alpha}D(i+j,t)\sum\limits_{\beta=\alpha+2}^{v_{max}}c_{\beta}^{\ast}(i,t_{1})}_\text{III}\\
&\quad +\underbrace{C(i+\alpha-1,t)D(i+\alpha,t)\prod_{j=1}^{\alpha-2}D(i+j,t)\sum\limits_{\beta=\alpha}^{v_{max}}c_{\beta}^{\ast}(i,t_{1})}_\text{IV}, \qquad 1<\alpha<v_{max}\\
c_{v_{max}}^{\ast}(i,t_{2})& =c_{v_{max}}^{\ast}(i,t_{1})\prod_{j=1}^{v_{max}}D(i+j,t)+C(i+v_{max}-1,t)  \\
&\quad
\times D(i+v_{max},t)\prod_{j=1}^{v_{max}-2}D(i+j,t)c_{v_{max}}^{\ast}(i,t_{1}).\\
\end{split}
\end{equation}
(iii) The braking stage: \\
\begin{equation}\label{NSOS2}
\begin{split}
c_{0}^{\ast}(i,t_{3})& =c_{0}^{\ast}(i,t_{2})+pc_{1}^{\ast}(i,t_{2}),\\
c_{1}^{\ast}(i,t_{3})& =\bar{p}c_{1}^{\ast}(i,t_{2})+p[c_{2}^{\ast}(i,t_{2})-C(i+1,t)D(i+2,t)\sum\limits_{\beta=2}^{v_{max}}c_{\beta}^{\ast}(i,t_{1})],\\
c_{\alpha}^{\ast}(i,t_{3})& =\underbrace{p[c_{\alpha+1}^{\ast}(i,t_{2})-C(i+\alpha,t)D(i+\alpha+1,t)\prod_{j=1}^{\alpha-1}D(i+j,t)\sum\limits_{\beta=\alpha+1}^{v_{max}}c_{\beta}^{\ast}(i,t_{1})]}_\text{I}\\
&\quad +\underbrace{\bar{p}[c_{\alpha}^{\ast}(i,t_{2})-C(i+\alpha-1,t)D(i+\alpha,t)\prod_{j=1}^{\alpha-2}D(i+j,t)\sum\limits_{\beta=\alpha}^{v_{max}}c_{\beta}^{\ast}(i,t_{1})]}_\text{II}\\ 
&\quad +\underbrace{C(i+\alpha-1,t)D(i+\alpha,t)\prod_{j=1}^{\alpha-2}D(i+j,t)\sum\limits_{\beta=\alpha}^{v_{max}}c_{\beta}^{\ast}(i,t_{1})}_\text{III}, \qquad 1<\alpha<v_{max}\\
c_{v_{max}}^{\ast}(i,t_{3})& =\bar{p}[c_{v_{max}}^{\ast}(i,t_{2})-C(i+v_{max}-1,t)D(i+v_{max},t)\prod_{j=1}^{v_{max}-2}D(i+j,t)c_{v_{max}}^{\ast}(i,t_{1})]\\
&\quad
+C(i+v_{max}-1,t)D(i+v_{max},t)\prod_{j=1}^{v_{max}-2}D(i+j,t)c_{v_{max}}^{\ast}(i,t_{1}).\\
\end{split}
\end{equation}
(iv) The motion stage:\\
\begin{equation}
\begin{split}
c_{\alpha}^{\ast}(i,t+1)=c_{\alpha}^{\ast}(i-\alpha,t_{3}).\qquad 0 \leq \alpha \leq v_{max}
\end{split}
\end{equation}

Since the overtaking vehicle is able to overtake the preceding one, it has more configurations than an ordinary vehicle. In the deceleration stage (ii), the item I describes that the overtaking vehicle at site $i$ with velocity $\alpha$ will keep its velocity if the preceding one do not locate in these $\alpha$ sites in the next time. If the preceding vehicle will locate at the site $i+\alpha+1$ in the next time, the vehicle with velocity $\alpha+1$ decreases one to avoid collision (item II). The NSOS model forbids to overtake more than one, so stage (ii) contains an item III. If an overtaking vehicle overtakes its preceding vehicle, it will close to the overtaken one according to our model, this contributes a situation with velocity $\alpha$ when a vehicle overtakes one with velocity $\alpha-1$ expressed by item IV. In the braking stage (iii), all the overtaking vehicles decrease by one with braking probability $p$ expect for successful overtaking vehicles, so one obtains an item I which the velocity with $\alpha+1$ of an overtaking vehicle minuses one with probability $p$ except for successful overtaking ones. Similarly, item II can be obtained without braking and the successful overtaking vehicle with velocity $\alpha$ also has a contribution to this stage (item III). 

Even though these time evolution equations are nonlinear, in the limit $t \rightarrow \infty $, the $C$ and $D$ distributions become homogeneous in space (for periodic boundary conditions). Therefore we can use the $C$'s values apart from the time and site dependences to calculate the traffic flow.

\begin{equation}\label{NSOS}
\begin{split}
c_{0}^{\ast}& =  pDc_{0}^{\ast}+pDCc_{1}^{\ast}+pDC^{2}\sum\limits_{\beta=2}^{v_{max}}c_{\beta}^{\ast},\\
c_{1}^{\ast} & = \bar{p}Dc_{0}^{\ast}+(\bar{p}DC+pD^{2})c_{1}^{\ast}+pD^{2}Cc_{2}^{\ast}+pD^{2}C^{2}\sum\limits_{\beta=3}^{v_{max}}c_{\beta}^{\ast}, \\
c_{\alpha}^{\ast} & = D^{\alpha-1}[(\bar{p}D+C)c_{\alpha-1}^{\ast}+(\bar{p}DC+pD^{2}+C)c_{\alpha}^{\ast}\\
&\quad +(\bar{p}DC+pD^{2}+1)Cc_{\alpha+1}^{\ast}+(\bar{p}DC+pD^{2}C+1)C\sum\limits_{\beta=\alpha+2}^{v_{max}}c_{\beta}^{\ast}], \qquad 1<\alpha<v_{max}\\
c_{v_{max}}^{\ast} & = D^{v_{max}-1}(\bar{p}D+C)(c_{v_{max}}^{\ast}+c_{v_{max}-1}^{\ast}).
\end{split}
\end{equation}

These equations are linear when we apply the relation $C=1-D$. So the equations (\ref{NS}) and (\ref{NSOS}) can be recast in matrix form as $\textbf{M}\vec{c}=\vec{c}$. The matrix $\textbf{M}$ can be read off from (\ref{NS}) and (\ref{NSOS}), $\vec{c}$ is the vector with elements $c_{\alpha}$, $\alpha=0,\dots,v_{max}$. For small $v_{max}$ one can calculate the probability $c_{\alpha}$ explicitly.

Since $c^{\dagger}=\bar{q}c$, $c^{\ast}=qc$ and $c=c^{\dagger}+c^{\ast}$, using the equations (\ref{NS}) and (\ref{NSOS}), we could obtain the specific form of $c_{\alpha}$ ($0 \leq \alpha \leq v_{max}$). Moreover, combined with the equation of flow $f(c,p,q)= \sum\limits_{\alpha=1}^{v_{max}}\alpha c_{\alpha}$, we can calculate the flow as a function of $p$, $q$ and $c$. Next, we calculate the flow in the case of $v_{max}=1$ and $v_{max}=2$, respectively. 

\subsection{Takeover: $v_{max}=1$}
No overtaking vehicles could overtake the preceding one in the limit of $v_{max}=1$. In fact, it becomes the takeover case which is discussed in the paper \cite{CHEN2001}, that is to say, overtaking vehicles could advance to the position that was occupied by their preceding ones at the previous time step. 

\begin{figure}[h]
	\centering
	\includegraphics[scale=0.4]{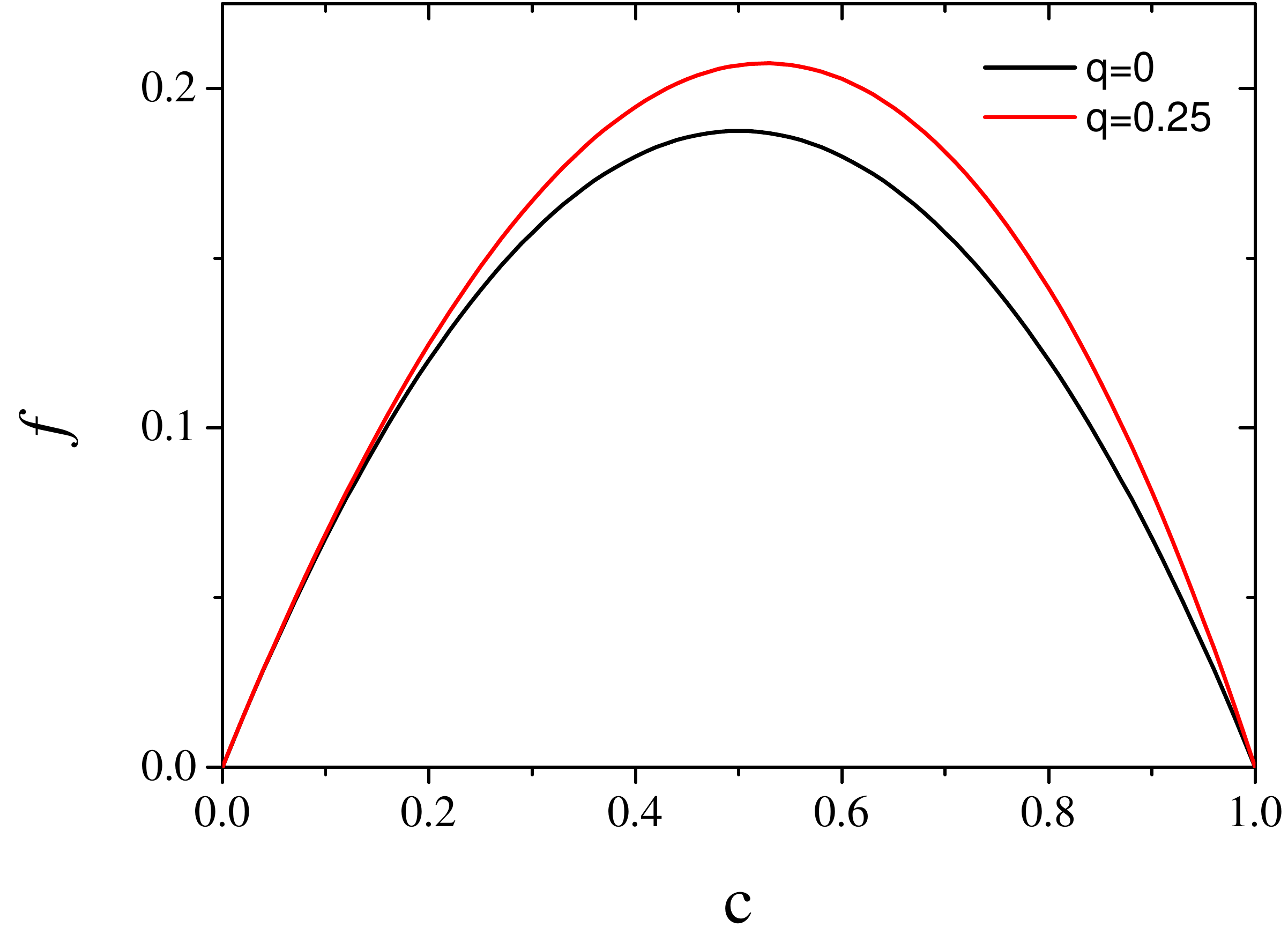}
	\caption{\label{mean1}(Color online) Fundamental diagram flow $f$ vs density $c$ for maximum velocity $v_{max}=1$ in the mean-field approximation in the case of $p=0.25$. Black line is in the case of $q=0$, and red line is in the case of $q=0.25$.}
\end{figure}
\begin{figure}[h]
	\centering
	\includegraphics[scale=0.4]{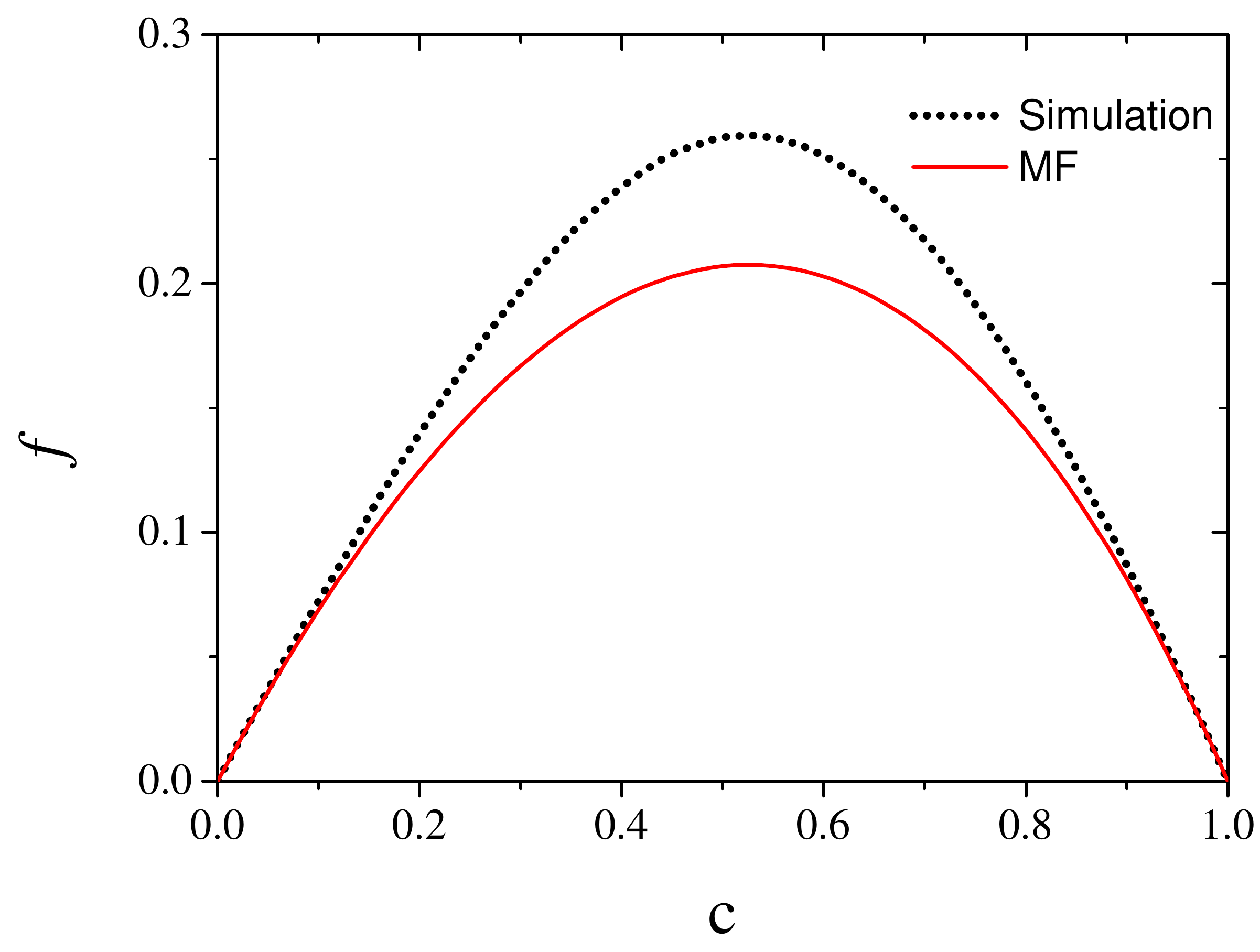}
	\caption{\label{mean2}(Color online) Fundamental diagram for $v_{max}=1$, $p=0.25$ and $q=0.25$. The red full curve is the MF result. For comparison the result from computer simulation (black dots) is also shown.}
\end{figure}
According to the update steps of the NSOS model, the time evolutions of these probability distributions can be described by the following four sets of equations:

(1) For the ordinary vehicles with $v_{max}=1$, the MF equations for the stationary state ($t \rightarrow \infty$) are given by \cite{Schreckenberg1995}
\begin{equation}
\begin{split}
c_{0}^{\dagger}& =(c+pd)c^{\dagger},\\
c_{1}^{\dagger}& =\bar{p}dc^{\dagger}.\\
\end{split}
\end{equation}
with $c^{\dagger}=c_{0}^{\dagger}+c_{1}^{\dagger}$.\\

(2) For the overtaking vehicles:

\begin{equation}
\begin{split}
c_{0}^{\ast}& =pDc^{\ast},\\
c_{1}^{\ast}& =\bar{p}Dc^{\ast}.\\
\end{split}
\end{equation}

Since $c^{\dagger}=\bar{q}c$, $c^{\ast}=pc$ and $c=c^{\dagger}+c^{\ast}$, we have

\begin{equation}\label{(V1)}
\begin{split}
c_{0}& =(c+pd)\bar{q}c+pqDc,\\
c_{1}& =\bar{p}\bar{q}dc+\bar{p}qDc.\\
\end{split}
\end{equation}

The flow $f(c,p,q)$ in the case $v_{max}=1$ is 

\begin{equation}
f(c,p,q)=c_{1}=\bar{p}\bar{q}dc+\bar{p}qDc.
\end{equation}
According to the definition of $D$, which denotes the empty possibility in the next time, considered the mean-field approximation it equals to the sum of distance $d$ of two neighborhood vehicles with the flow $f$, i.e. $D \approx d+f$. So the equation becomes
\begin{equation}
f(c,p,q)=\frac{\bar{p}dc}{1-\bar{p}qc}.\\
\end{equation}

The first information from this equation is that the flow is dominated by braking probability $p$, while the overtaking probability $q$ is not the important factor since $q$ has a factor $\bar{p}c$. Specially, the flow becomes the form of NS model when $q=0$. Another finding is that the flow is enhanced mainly in the jammed phase. If we keep the $p$ and $q$ invariant, with the growth of density $c$, the denominator decreases monotonically, while the value of flow increases larger. This result can be explained in physical terms. In the free flow phase, all the vehicles move freely which means the preceding ones has no influence to overtaking vehicles, the impact of $D$ on the flow is the same as the one of $d$. While in the jammed phase, overtaking vehicles have larger probabilities to move than ordinary ones due to $D>d$, so the flow in this phase is larger than the one of NS model. Moreover, the larger $q$ induces, the quicker flow increases.

The mean-field result yields, compared with the simulation data shown in Fig.\,\ref{mean2}, much too small values of the flow. This can easily be understood since the reduction to a single vehicles problem ignores all spatial correlations of the vehicles \cite{Schreckenberg1995}.       

\subsection{Overtaking: $v_{max}=2$}
It is the simplest case that overtaking vehicles could overtake the preceding ones in the NSOS model in the case of $v_{max}=2$. Moreover, it is natural without innateness hypothesis that an overtaking vehicle is only able to overtake one each time and occupy the site just in front of its preceding one if it could overtake successfully. So it is necessary to calculate the exact equations in the case of $v_{max}=2$. Using the (\ref{NS})-(\ref{NSOS2}) equations, we have:

\begin{figure}[h]
	\centering
	\includegraphics[scale=0.4]{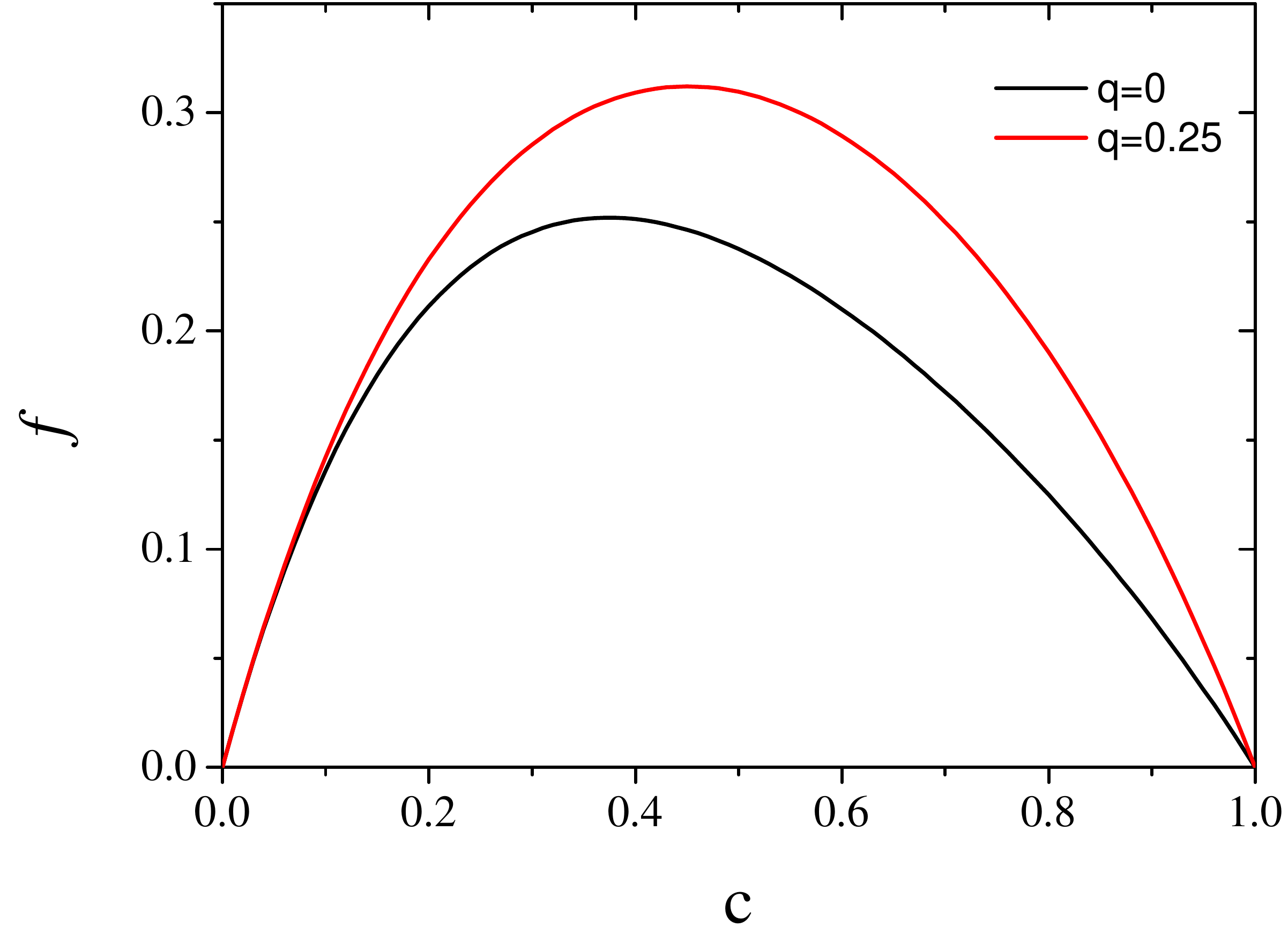}
	\caption{\label{mean3}(Color online) Fundamental diagram flow $f$ vs. density $c$ for maximum velocity $v_{max}=2$ in the mean-field approximation in the case of $p=0.25$. Black line is in the case of $q=0$, and red line is in the case of $q=0.25$.}
\end{figure}
\begin{figure}[h]
	\centering
	\includegraphics[scale=0.4]{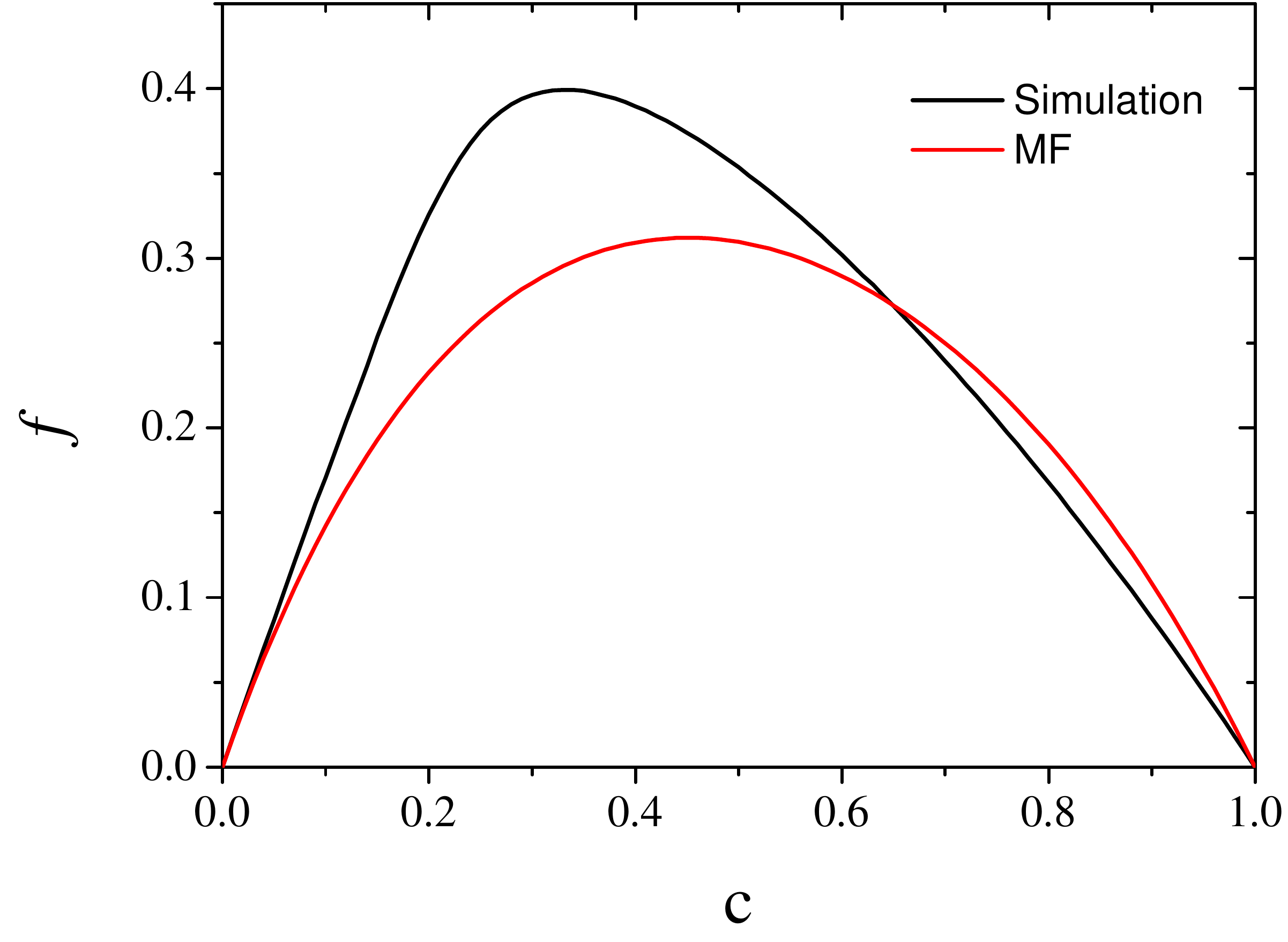}
	\caption{\label{mean4}(Color online) Fundamental diagram for $v_{max}=2$, $p=0.25$ and $q=0.25$. The  red full curve is the MF result. For comparison the result from computer simulation (black dots) is also shown.}
\end{figure}

(1) For the ordinary vehicles:

\begin{equation}
\begin{split}
c_{0}^{\dagger}& =\frac{(1+pd)c}{1-pd^{2}}c^{\dagger},\\
c_{1}^{\dagger}& =\frac{\bar{p}(1-\bar{p}d^{2})d}{1-pd^{2}}c^{\dagger},\\
c_{2}^{\dagger}& =\frac{\bar{p}^{2}d^{3}}{1-pd^{2}}c^{\dagger}.
\end{split}
\end{equation}

(2) For the overtaking vehicles:

\begin{equation}\label{V2}
\begin{split}
c_{0}^{\ast}& =\frac{pDC}{1-pD^{2}}c^{\ast},\\
c_{1}^{\ast}& =\frac{\bar{p}C+pD(1-pD)}{1-pD^{2}}Dc^{\ast},\\
c_{2}^{\ast}& =\frac{(\bar{p}D+C)(1-pD)}{1-pD^{2}}Dc^{\ast}.\\
\end{split}
\end{equation}
Here, we have used equations $C+D=1$ and $c_{1}^{\ast}+c_{2}^{\ast}=c^{\ast}-c_{0}^{\ast}$. Again, we could calculate $c_{\alpha}$ ($\alpha=0,1,2$) using $c_{\alpha}=c_{\alpha}^{\dagger}+c_{\alpha}^{\ast}$ and $c^{\dagger}=\bar{q}c$, $c^{\ast}=qc$. But this time we assume $D \approx d$ due to the value of $D-d$ is smaller than $d$. The flow can be calculated using the following equation $f(c,p,q)=(c_{1}^{\dagger}+c_{1}^{\ast})+2(c_{2}^{\dagger}+c_{2}^{\ast})$. The result is shown in Fig.\,\ref{mean3}. One could also observe that the flow of the NSOS model enlarged in the jammed regime than that of original NS model, this would be due to overtaking mechanism is beneficial to develop the traffic flow. Again, the mean-filed result is still less than simulation data (Fig.\,\ref{mean4}), and one could observe that the maximum flow density does not coincide with the simulation result, this may be the result of our simplicity $D$.   

\section{Conclusions}
In this paper theoretical analysis of the NSOS model is performed by using the mean-field method, the equations for $v_{max}=1$ can be obtain exactly, while for larger values of $v_{max}$ they are just approximations. Even though mean-field theory is insufficient due to the important correlations between neighboring sites are neglected, the reason that why the NSOS model can improve traffic flow in the area where the flow exceed the maximum flow density has been explained, and braking probability as a major factor that influence transition density has been discovered. 

\section{Acknowledgments}
This work was supported by China Postdoctoral Science Foundation (Grant Nos. 30205010003), Fundamental Research Funds for the Central Universities (Grant Nos. 20205170444), and National Natural Science Foundation of China (Grant Nos. 11505071 and 61702207).

\section*{Appendix A: Takeover Case}
In this appendix we show that the stationary state of the NSOS model with $v_{max}=1$ for overtaking vehicles:

(i) The acceleration stage:

\begin{equation}
\begin{split}
c_{0}^{\ast}(i,t_{1})& =0,\\
c_{1}^{\ast}(i,t_{1})& =c_{1}^{\ast}(i,t)+c_{0}^{\ast}(i,t).\\
\end{split}
\end{equation}

(ii) The deceleration stage:

\begin{equation}
\begin{split}
c_{0}^{\ast}(i,t_{2})& =c_{0}^{\ast}(i,t_{1}),\\
c_{1}^{\ast}(i,t_{2})& =D(i+1,t)c_{1}^{\ast}(i,t_{1}).\\
\end{split}
\end{equation}

(iii) The braking stage:

\begin{equation}
\begin{split}
c_{0}^{\ast}(i,t_{3})& =c_{0}^{\ast}(i,t_{2})+pc_{1}^{\ast}(i,t_{2}),\\
c_{1}^{\ast}(i,t_{3})& =\bar{p}c_{1}^{\ast}(i,t_{2}).\\
\end{split}
\end{equation}

(iv) The motion stage:

\begin{equation}
\begin{split}
c_{\alpha}^{\ast}(i,t+1)& =c_{\alpha}^{\ast}(i-\alpha,t_{3}), \qquad 0 \leq \alpha \leq 1.
\end{split}
\end{equation}

In the stationary state, distributions become homogeneous in space for periodic boundary conditions, so the site dependence could be omitted. Using this and combining the four update steps one gets the set of equations (\ref{(V1)}).

\section*{Appendix B: Overtaking Case}
In this appendix we show that the stationary state of the NSOS model with $v_{max}=2$ for overtaking vehicles:

(i) The acceleration stage:

\begin{equation}
\begin{split}
c_{0}^{\ast}(i,t_{1}) & = 0,\\
c_{1}^{\ast}(i,t_{1}) & = c_{0}^{\ast}(i,t),\\
c_{2}^{\ast}(i,t_{1}) & = c_{2}^{\ast}(i,t)+c_{1}^{\ast}(i,t).\\
\end{split}
\end{equation}

(ii) The deceleration stage:

\begin{equation}
\begin{split}
c_{0}^{\ast}(i,t_{2})& =c_{0}^{\ast}(i,t_{1}),\\
c_{1}^{\ast}(i,t_{2})& =D(i+1,t)c_{1}^{\ast}(i,t_{1})+D(i+1,t)C(i+2,t)c_{2}^{\ast}(i,t_{1}),\\
c_{2}^{\ast}(i,t_{2})& =D(i+1,t)D(i+2,t)c_{2}^{\ast}(i,t_{2})+C(i+1,t)D(i+2,t)c_{2}^{\ast}(i,t_{1}).\\
\end{split}
\end{equation}

(iii) The braking stage:

\begin{equation}
\begin{split}
c_{0}^{\ast}(i,t_{3})& =c_{0}^{\ast}(i,t_{2})+pc_{1}^{\ast}(i,t_{2}),\\
c_{1}^{\ast}(i,t_{3})& =\bar{p}c_{1}^{\ast}(i,t_{2})+p[c_{2}^{\ast}(i,t_{2})-C(i+1,t)D(i+2,t)c_{2}^{\ast}(i,t_{1})],\\
c_{2}^{\ast}(i,t_{3})& = \bar{p}[c_{2}^{\ast}(i,t_2)-C(i+1,t)D(i+2,t)c_{2}^{\ast}(i,t_{1})]+C(i+1,t)D(i+2,t)c_{2}^{\ast}(i,t_{1}).\\
\end{split}
\end{equation}

(iv) The motion stage:

\begin{equation}
\begin{split}
c_{\alpha}^{\ast}(i,t+1)& =c_{\alpha}^{\ast}(i-\alpha,t_{3}), \qquad 0 \leq \alpha \leq 1.
\end{split}
\end{equation}

In the stationary state, distributions become homogeneous in space for periodic boundary conditions, so the site dependence could be omitted. Using this and combining the four update steps one gets the set of equations (\ref{V2}).

\end{document}